\documentclass[copyright,creativecommons]{eptcs}
\providecommand{\event}{ML Workshop 2015} 

\usepackage{breakurl}             
\usepackage{amsmath}
\usepackage{listings}
\usepackage{amsfonts}
\usepackage{amsthm}
\usepackage{courier}
\usepackage{fancyvrb}
\usepackage[frozencache=true]{minted}

\title%
{%
Dependent Types for Multi-Rate Flows\break in Synchronous
Programming\break (System Description)%
}%
\author{William Blair
\institute{Boston University\\ Massachusetts, USA}
\email{wdblair@cs.bu.edu}
\and
Hongwei Xi
\institute{Boston University\\
Massachusetts, USA}
\email{hwxi@cs.bu.edu}
}
\def\titlerunning%
{Dependent Types for Multi-Rate Flows}
\def\authorrunning{William Blair \& Hongwei Xi}
\begin{document}
\maketitle

\begin{abstract}
Synchronous programming languages emerged in the 1980s as tools for
implementing reactive systems, which interact with events from
physical environments and often must do so under strict timing
constraints. In this report, we encode inside ATS various real-time
primitives in an experimental synchronous language called Prelude,
where ATS is a statically typed language with an ML-like functional
core that supports both dependent types (of DML-style) and linear
types. We show that the verification requirements imposed on these
primitives can be formally expressed in terms of dependent types in
ATS. Moreover, we modify the Prelude compiler to automatically
generate ATS code from Prelude source. This modified compiler allows
us to solely rely on typechecking in ATS to discharge proof
obligations originating from the need to typecheck Prelude code.
Whereas ATS is typically used as a general purpose programming
language, we hereby demonstrate that it can also be conveniently used
to support some forms of advanced static checking in languages
equipped with less expressive types.
\end{abstract}

\section%
{Introduction}
Building software that must work reliably is an undeniably challenging
process. Yet, engineers can still construct reliable real-time systems
that operate under strict temporal requirements. Companies have long
adopted model-based methods for designing real-time
software\cite{Benveniste2003}. Model-based design methods emphasize
automatically generating code from high-level models that formally
describe the behavior of a system. In fact, synchronous programming
languages helped start a movement towards this paradigm (of
model-based automatic code generation) when they were introduced in
the 1980s.

In synchronous programs, tasks communicate through data flows with the
assumption that each data flow produces values instantaneously at
every step of a global logical clock. However, the simplicity of a
single system-wide global logical clock can be the source of serious
limitation in practice.  If a system consists of communicating
periodic tasks that need to run at different rates, the global logical
clock can become an obstacle as the programmer must synchronize all
the tasks manually. The process of manual synchronization is likely to
be both tedious and error-prone with little support provided by the
compiler for preventing subtle timing bugs. As a part of his PhD
thesis, Julien Forget developed Prelude\cite{Forget09}, a synchronous
language that features primitives for describing the real-time
behavior of tasks in a multi-rate periodic system. Synchronizing data
flows with different clocks in Prelude is still left to the
programmer, but the language provides operators to directly modify
real-time clocks. For example, a programmer can stretch a clock to
accomplish under-sampling or shrink it for over-sampling. The
typechecker of Prelude then automatically verifies that all
communicating tasks are synchronized. The advantage of this system is
that the language offers the ability to easily adjust the temporal
behavior of tasks while providing support for verification in the type
system.

In this report, we encode inside ATS the clock calculus presented in
Prelude, where ATS is a statically typed programming language with a
functional core of ML-style. The type system of ATS is rooted in the
Applied Type System framework\cite{ATS-types03}, which gives the
language its name. In ATS, both dependent types\cite{XP99,DML-jfp07}
(of DML-style) and linear types are supported. We assign dependent
types to the real-time clocks available in Prelude and then develop
functions to capture the behavior of clock transformation operators.
With these functions, we can typecheck the automatically generated ATS
code from Prelude source (by a modified Prelude compiler) and formally
verify that the data flows in the original Prelude source are properly
synchronized. A running version of our modified compiler can be found
online\footnote{An end-to-end demonstration of our system is
  available at \url{https://travis-ci.org/wdblair/overture}}.

Although ATS is typically used as a source programming language, we
hereby demonstrate that it can also be conveniently used as a target
language for the purpose of supporting advanced static checking in a
host language equipped with less expressive types. In addition, the
reported use of dependent types for reasoning about temporal
properties should be interesting in its own regard.

\section%
{Multi-Rate Flows in Prelude}\label{section:Prelude}
An embedded real-time system continuously interacts with its own
environment and often must do so under strict timing constraints. For
example, a task may periodically sample a sensor and provide data that
can in turn be used by the application logic to adjust the output of
an actuator. In order to guarantee that the system is responsive,
engineers need to demonstrate that the worst case execution time
(WCET) for the communication from the sensor, to the application
logic, and then to the actuator can never exceed a required
deadline. Synchronous programming languages simplify development by
abstracting time for the designer and providing well-defined semantics
to enable verification and automatic code generation from a model of
the system.

When a synchronous program is compiled, all of the nodes and flows are
translated into a single-step function that is invoked repeatedly. The
time between steps is the period of the system, and the Synchronous
Hypothesis states that all the flows must be re-evaluated before the
end of each step. It is up to the system designer to ensure that this
requirement is met. Several synchronous
languages\cite{Benveniste1991,Berry1992} have been developed since the
1980s, with Lustre\cite{Pilaud1987} currently serving as the
underlying language powering the SCADE Suite developed by Esterel
Systems.

This abstraction for time in a synchronous language is of great
convenience because it not only alleviates the need for the programmer to
worry about low-level timing details but also produces fully
deterministic code. However, it makes it difficult at the same time to
compose systems where tasks may run at different rates.  In order to
make tasks execute at the desired rates, the programmer must
essentially schedule tasks by hand in terms of a global logical clock,
which can be both tedious and error-prone.

Prelude addresses the problem of manual synchronization for multi-rate
periodic systems by removing the notion of a single system-wide global
clock. Instead, it assigns each periodic flow $F$ its own clock $C$
represented as a pair $(n,p)$ where $n\in\mathbb{N}^+$ (that is, $n$
is a positive integer) and $p\in\mathbb{Q}^+$ (that is, $p$ is a
positive rational number); the integer $n$ is referred to as the
period of $C$ and $p$ the phase offset. Note that the period of a flow
$F$ is just the inter-arrival time of values produced by $F$ and the
phase offset is the initial delay of the flow. The activation time of
a flow (that is, its start date) is given by $n \cdot p$. In Prelude's
clock calculus, every date (that is, a point in time in the program)
is required to be an integer so as to simplify schedulability
analysis. Therefore, we need to ensure that the date $t_0$ is an
integer for every flow.  In other words, a clock defined as $(n,p)$ is
valid if and only if both $n\in\mathbb{N}^{+}$ and $n \cdot
p\in\mathbb{N}^{+}$ hold.

In Prelude, a generic flow consists of an infinite list of tuples
$\forall i\in\mathbb{N}.~(v_i, t_i)$, where $v_i$ is the value of the
flow produced at time $t_i$. Under the normal synchronous paradigm,
every flow produces some value during each tick of the logical
clock. There is no longer the notion of a global time scale in
Prelude. Instead, every flow is assigned its own clock, which can be
lengthened as well as shortened. By modifying a clock, we alter the
rate at which values are produced in a program. If a value contained
in a flow is undefined for some arbitrary date, we cannot adjust its
real-time behavior while expecting deterministic results.  As such, it
is useful to define a class of data flows that are strictly
periodic. Formally speaking, a data flow is strictly periodic if and
only if
$$\exists n \in \mathbb{N}^{+}. \forall i \in \mathbb{N}.
  \text{  } t_{i+1} - t_{i} = n$$
where $n$ is the period of the clock assigned to the flow and $t_{i}$
is the date when value $i$ is produced by the flow\cite{Forget09}.
Intuitively, a strictly periodic data flow produces a value after each
period. In Prelude, a set of operators are provided to transform the
clocks assigned to strictly periodic flows which greatly facilitates
communication between tasks that execute at different rates. In
Figure~\ref{figure:clock_trans_opers} presented in
Section~\ref{sec:atstypes} we describe how to encode the semantics of
these operators into the ATS programming language.

\begin{figure}
\begin{minted}{prelude}
imported node
database (i: int rate (10, 0))
  returns (o: int rate (10, 0));

imported node
controller(i: int rate (100, 0); j: int rate (100, 0))
  returns (o: int rate (100, 0);  p: int rate (100, 0));

sensor i;
actuator o;

node
sampling (i: rate (10, 0)) returns (o: rate (100, 0))
  var command: rate (100, 0);
  var response: rate (10, 0);
let
  (o, command) = controller(i/^10, (0 fby response)/^10);
  response = database(command*^10);
tel
\end{minted}
\caption{A multi-rate periodic system in the Prelude language}
\label{figure:Prelude_example}
\end{figure}
As an example, let us consider a simple system where a controller runs
with a clock of $(100,0)$ and interacts with a fast running task that
provides critical data from the environment with a clock of $(10,0)$.
Figure~\ref{figure:Prelude_example} implements this system where the
\textbf{controller} node sends a command to the \textbf{database} node
and receives a response back from it. Observe that the
\textbf{controller} executes 10 times slower than the
\textbf{database}. If we want the \textbf{controller} node's output
\textbf{command} to be synchronized with the \textbf{database}, we
must over-sample it using the \mintinline{prelude}{*^} operator.
Likewise, we under-sample the \textbf{database} node's output
\textbf{response} in order to synchronize it with the
\textbf{controller} node by using the \mintinline{prelude}{/^}
operator.

In the rest of the paper, we will define the Prelude operators
necessary to implement this program in ATS so that all synchronization
constraints are checked by the type checker. Then, we will describe
how we can transform Prelude programs like this one into ATS to verify
all communication is synchronized.

\section%
{Assigning Types to Flows in ATS}
\label{sec:atstypes}
At a high level, Prelude modifies the semantics of traditional
synchronous languages by refining the basic flow type to have its own
clock as opposed to a global clock. This refinement generates a proof
obligation for the compiler as a flow going from one task to
another is not necessarily synchronized. We can readily
defer this obligation to typechecking in ATS by assigning the following
type (written in the concrete syntax of ATS) to strictly periodic flows:

\def\snat{\mbox{\it nat}}
\def\sint{\mbox{\it int}}
\def\srat{\mbox{\it rat}}
\def\stype{\mbox{\it type}}
\def\tint{\mbox{\bf int}}
\def\tbool{\mbox{\bf bool}}
\def\tSFlow{\mbox{\bf SFlow}}
\def\tBFlow{\mbox{\bf BFlow}}
\begin%
{minted}{sml}
abstype SFlow(a: type, n: int, p: rat)
\end{minted}
Note that $\sint$ and $\srat$ refer to the sorts for integers and
rational numbers, respectively. There is another sort $\snat$ for
natural numbers that will appear later.  Given $a$, $n$, and $p$, the
type $\tSFlow(a, n, p)$ is for a strictly periodic flow of values of
the type $a$ such that the clock assigned to the flow is of the period
$n$ and phase offset $p$. Formally speaking, a flow $F$ of the type
$\tSFlow(a, n, p)$ consists of the following pairs:
$$\forall i \in \mathbb{N}. (v_i, t_i) \in F$$
where each $v_i$ is a (defined) value of the type $a$ and $t_i =
n\cdot(p + i)$.  Flows that are not strictly periodic are also useful
in Prelude. For instance, the following type $\tBFlow$ is for boolean
flows:
\begin%
{minted}{sml}
abstype BFlow(a: type, n: int, p: rat)
\end{minted}
Boolean flows differ from strictly periodic flows by allowing values
to be produced only on a subset of the dates given by a flow's clock.
In Figure~\ref{figure:Prelude_bool_example}, we define a boolean flow
\mintinline{prelude}{alert} to give the value of a temperature sensor
only at dates when \mintinline{prelude}{fault} produces a
\mintinline{prelude}{true} value.

\begin{figure}
\begin{minted}{prelude}
node
monitor (temperature: int rate (10, 0), fault: bool rate(100, 0))
  returns (alert: int rate (100, 0))
let
  alert = (temperature /^ 10) when fault
tel
\end{minted}
\caption{The boolean flow \mintinline{prelude}{alert} given here
  produces data only when \mintinline{prelude}{fault} produces
  \mintinline{prelude}{true}.}
\label{figure:Prelude_bool_example}
\end{figure}

Modifying a flow's clock affects the values that are included in the
flow. In order to determine precisely what values remain in a flow
after its clock is modified, Prelude requires that every flow passed
to a clock transformation operator be strictly periodic. By the same
token, all clock operators yield strictly periodic flows.

In Prelude, only integer dates are considered valid. Therefore, any
flow's clock is subject to the requirement that $n \cdot p \in
\mathbb{N}$ holds, where $n$ and $p$ are the period and phase offset
of the clock, respectively. If this was not true, then a task's start
time could not be represented as an integer date. When assigning types
to a clock transformation operator, we often add a precondition to
ensure that only a clock satisfying the requirement can be actually
returned by the operator.

\def\fcons{\mbox{\tt cons}}
\def\ftail{\mbox{\tt tail}}
The clock transformation operators in Prelude are to be translated
into regular functions in ATS.  Synchronous flows resemble infinite
lazy streams in traditional functional programming languages. For
instance, we can introduce an operator $\fcons$ of the following type:
\[
\begin%
{array}{rcl}
$\fcons$ &:& \forall a:\stype\,\forall n:\sint\,\forall p:\srat.~(a, \tSFlow(a,n,p)) \rightarrow \tSFlow(a, n, p-1)\\
\end{array}\]
Basically, $\fcons$ adds a given value to the beginning of a given
flow to form a new flow that starts one period ahead of the given
flow; the first produced value of the new flow is simply the given
value, and the rest are those produced by the given flow. As another
example, we can introduce an operator $\ftail$ of the following type:
\[
\begin%
{array}{rcl}
$\ftail$ &:& \forall a:\stype\,\forall n:\sint\,\forall p:\srat.~\tSFlow(a,n,p) \rightarrow \tSFlow(a, n, p+1) \\
\end{array}\]
Given a flow, $\ftail$ returns another one
that simply delays the given flow's activation by one period.

\def\fmul{\mbox{\tt mul}}
\def\fdiv{\mbox{\tt div}}
\def\ffby{\mbox{\tt fby}}
\def\fshift{\mbox{\tt shift}}
\begin%
{figure}
\[%
\begin%
{array}{rcl}
\fmul &\kern-3pt:\kern-3pt& \forall a:\stype\,\forall n:\sint\,\forall p:\srat\,\forall k:\snat.~
(k \mid n)\supset(\tSFlow(a,n,p), \textbf{int} (k)) \rightarrow \tSFlow (a, n / k, p \cdot k)\\
\fdiv &\kern-3pt:\kern-3pt& \forall a:\stype\,\forall n:\sint\,\forall p:\srat\,\forall k:\snat.~
(k \mid n)\supset(\tSFlow(a,n,p), \tint(k)) \rightarrow \tSFlow (a, n \cdot k, p / k)\\
\fshift &\kern-3pt:\kern-3pt& \forall a:\stype\,\forall n:\sint\,\forall p:\srat\,\forall k:\srat.~
(n \cdot k\in\mathbb{N}) \supset \ (\tSFlow(a,n,p), \textbf{rat}(k)) \rightarrow \tSFlow (a, n, p+k)\\
\ffby &\kern-3pt:\kern-3pt& \forall a:\stype\,\forall n:\sint\,\forall p:\srat.~(a, \tSFlow (a,n,p)) \rightarrow \tSFlow (a, n, p)\\
\end{array}\]
\label{figure:clock_trans_opers}
\caption{Some clock transformation operators and their types}
\end{figure}
There are four commonly used clock transformation operators in
Figure~\ref{figure:clock_trans_opers} plus the types assigned to them.
The $\fmul$ operator (that is, \verb|*^| in Prelude)
speeds up a clock by shrinking its period; it implements over-sampling
where every element in the flow is repeated $k$ times to meet its
timing constraint. Likewise, the $\fdiv$ operator (that is, \verb|/^|
in Prelude) slows a clock by stretching its period; it implements
under-sampling where only one element is taken out of every $k$
elements of the input flow.  Given a flow and a rational number $k$,
the $\fshift$ operator shifts the phase offset of the clock assigned
to the flow by $k$, thus delaying the activation of the flow. Given an
element and a flow, the $\ffby$ operator does not modify the flow's
clock at all; it instead simply shifts the flow by one period and sets
the first element of the flow with the given one. For instance,
$\ffby$ can be implemented based on $\fcons$ and $\fshift$ in the
concrete syntax of ATS as follows:
\begin%
{minted}{sml}
implement fby (x, f) = cons(x, shift(f, 1))
\end{minted}

\def\fwhen{\mbox{\tt when}}
\def\fmerge{\mbox{\tt merge}}
Another interesting operator $\fmerge$ is given the following type:
$$%
\forall a:\stype\,\forall n:\sint\,\forall p:\srat.~%
(\tSFlow(\tbool,n,p), \tSFlow(a,n,p), \tSFlow(a,n,p)) \rightarrow \tSFlow(a, n, p)%
$$%
where $\tbool$ is the type for booleans. Essentially, $\fmerge$ takes
three strictly periodic flows and returns one; for each $i$, if the
value in the first flow at $t_i$ is true, then the value in the second
flow at $t_i$ is picked to be the value of the returned flow at $t_i$;
otherwise, the value in the third flow at $t_i$ is picked.

Yet another interesting operator $\fwhen$ is given the following type:
$$%
\forall a:\stype\,\forall n:\sint\,\forall p:\srat.~
(\tSFlow(\tbool,n,p), \tSFlow(a,n,p)) \rightarrow \tBFlow(a,n,p)
$$%
where $\tBFlow$ is for boolean flows. Given two strictly periodic
flows, $\fwhen$ returns a boolean flow; for each $i$, if the value of
the first flow is true at $t_i$, then the value of the second flow at
$t_i$ is picked to be the value of the returned flow at $t_i$;
otherwise, the returned flow is undefined at $t_i$. Note that $\fwhen$
is so far the only operator that does not return a strictly periodic
flow.

So far we have only thought of communication through flows in Prelude
as individual values arriving at dates (that is, points in
time). Please recall the simple example presented in
Figure~\ref{figure:Prelude_example}, where a controller queries a faster data
task for information.  In order to make the communication between them
synchronous, we specified that only one out of every 10 values
produced by the faster task is to be read by the controller.

\def\fdivque{\mbox{\tt div\_queue}}

Suppose that the controller needs all values produced by the
database since it was last invoked. This requires a queuing mechanism
in the clock calculus, which is a feature outlined in Prelude's
specification but yet to be supported by the compiler.  We can readily
add verification support for this feature with dependent types in
ATS. Suppose we have a strictly periodic flow $F$ that we want to
under-sample by some factor $k$. We can think of the flow as having
values of the type $array(a:type, k:int)$ for arrays that each
contain $k$ elements of the type $a$. This gives us the following
type for a divide operation that performs queuing rather than
under-sampling. In this way, every time the node that consumes this
flow is activated, it has access to all $k$ of the most recent values
emitted from the flow. Putting this formally, the type for such an operator
is as follows:
$$%
\forall a:\stype\,\forall n:\sint\,\forall p:\srat\,\forall k:\sint.~%
 (k > 0)\supset(\tSFlow(a,n,p), \tint(k)) \rightarrow \tSFlow(array(a,k), n\cdot k, p/k)
$$

Clearly, this type directly relates the number of accumulated values
produced by a given flow to the factor by which the flow's clock is
expected to be stretched.

\section%
{Using ATS as a Target Language}

With the above outlined operators plus their types, we can write ATS
programs that correspond semantically to their Prelude counterparts in
terms of the synchronization checks that must be performed. In this
work, we modified the Prelude compiler to automatically generate ATS
code with type signatures from Prelude source. As we will see in this
section, the translation is largely straightforward.

Every imported node is declared as an external function with their
input and output flows having the same clock, and every node that is
actually implemented in Prelude is translated into a function
definition in ATS. Probably the biggest issue is to bridge the
semantic gap between Prelude, a concurrent declarative language, and
ATS, a call-by-value language with an ML-like functional core. This is
especially evident when we translate programs where certain flows are
passed as arguments before they are actually defined. For instance,
please take a careful look at the code presented in
Figure~\ref{figure:Prelude_example}. This issue cannot be addressed by
simply reordering expressions as circular definitions are both legal
and common in Prelude. Instead, we address this issue by using a
combination of linear types and support for proof terms in ATS. For
each locally declared flow, we generate a linear proof term. In order
to consume this proof term, we specify a proof function our Prelude
compiler must call with two ATS flows that have equal clocks to ensure
the Prelude flow they represent is given exactly one clock. In
Figure~\ref{figure:ATS_translation} you can see this approach used to
verify the \mintinline{prelude}{response} and
\mintinline{prelude}{command} flows maintain their specified clock.

\begin{figure}
\begin{minted}{sml}
fun main (
  i: SFlow (int, 10, 0)
): (SFlow (int, 100, 0)) = let
  var response : SFlow (int, 10, 0)
  prval pfresponse = flow_future_make (response)
  var command : SFlow (int, 100, 0)
  prval pfcommand = flow_future_make (command)
  val response' = (flow_fby (0, response))
  val o, command' = controller ((flow_div_clock (i, 10)),
                                  (flow_div_clock (response', 10)))
  val response' = server (flow_mul_clock (command, 10))
  prval () = flow_future_elim (pfresponse, response, response')
  prval () = flow_future_elim (pfcommand, command, command')
in
  (o)
end
\end{minted}
\caption%
[ATS code automatically generated by a modified Prelude compiler]%
 {ATS code automatically generated by a modified Prelude
   compiler}
\label{figure:ATS_translation}
\end{figure}
After modifying the Prelude compiler, we automatically generate ATS
code directly from the abstract syntax tree of Prelude source;
potential synchronization errors in the Prelude source can be captured
by typechecking the ATS code translated from it. For instance, the
Prelude example in Figure~\ref{figure:Prelude_example} is translated
into the ATS program in Figure~\ref{figure:ATS_translation} for the
purpose of typechecking. Note that the lines starting with the keyword
{\tt prval} are theorem-proving code related to the circularly defined
flows in this function.

\section{Conclusion and Future Work}

Prelude simplifies the process of composing communicating real-time
tasks, that possibly run at different rates, into a fully
deterministic system. This desire for rooting out nondeterminism in
the development process was present when synchronous languages were
first introduced\cite{halbwachs1992} and it continues with methods
applied for industrial tools like Matlab's Simulink\cite{clarke2011}
and Esterel Technologie's SCADE which uses the synchronous language
Lustre in mission critical software across aerospace, automotive, and
industrial applications.

There is also research on synchronous languages that addresses the
implementation of Mixed Criticality Systems\cite{yip2014}, where some
tasks may be allowed to violate the Synchronous Hypothesis and miss
deadlines. One could imagine extending a language like Prelude to
guarantee synchronized deterministic communication between both hard
real-time tasks and less critical soft real-time tasks. The Liquid
Clocks\cite{talpin2015towards} framework defines a type system for
designing new synchronous languages across different models of time
and communication. Liquid Clocks uses refinement types and an
inference algorithm inspired by Liquid Types\cite{rondon2008liquid} to
verify time and causality constraints in a network defined by data
flows. Our approach differs from Liquid Clocks in that we use the
expressive types found in ATS, an existing programming language, to
support the implementation of a domain specific synchronous language
for multi-rate periodic systems.

In this report, dependent types in ATS of
DML-style~\cite{XP99,DML-jfp07} are assigned to the multi-rate flows
introduced in Prelude and also to various clock transformation
operators on these flows. Note that there is explicit use of
quantifiers in the types assigned to these operators, far exceeding
what is accomplished in Prelude, where every clock is either fixed to
be a constant or can be inferred from a constant one. For instance, a
constraint stating that a node must always produce a flow whose clock
is twice as fast as its input node can be readily expressed in ATS but
cannot currently in Prelude.

The basic building block of composition in synchronous languages like
Prelude is a unidirectional flow. Yet, in complex embedded systems,
individual components may communicate using various different
protocols. While these protocols are implemented on top of flows,
there is no formal description for them in an architecture design
language like Prelude. One avenue for future work is to develop a
synchronous domain specific version of ATS that can use typed channels
like those found in Session Types\cite{honda1993types} to describe
both the temporal behavior of the system and what is communicated
inside of it. The latter information, for instance, can help verify
node implementations in a synchronous program or detect possible
errors.

\section{Bibliography}

\nocite{*}
\bibliographystyle{eptcs}
\bibliography{dependent-types-mrate-flows}
\end{document}